# Nanostructured ZnO films: a study of molecular influence on transport properties by impedance spectroscopy

Luciano D. Sappia [a,b]‡, Matias R. Trujillo [a,b]‡, Israel Lorite [d], *Rossana E. Madrid [a,b], Monica Tirado [c], David Comedi [c], **Pablo Esquinazi [d]

[a] Instituto Superior de Investigaciones Biológicas (INSIBIO), CONICET, Chacabuco 461, T4000ILI – San Miguel de Tucumán, Argentina.

[b] Laboratorio de Medios e Interfases (LAMEIN), Departamento de Bioingeniería, Fac. de Cs. Exactas y Tecnología, Universidad Nacional de Tucumán, Av. Independencia 1800, 4000 – San Miguel de Tucumán, Argentina.

[c] Laboratorio de Nanomateriales y Propiedades dieléctricas de la Materia, Departamento de Física, Universidad Nacional de Tucumán, Avenida Independencia 1800, Tucumán, Argentina.

[d] Superconductivity and Magnetism Division – University of Leipzig - Leipzig, Germany.

Corresponding authors: *rmadrid@herrera.unt.edu.ar; **esquin@physik.uni-leipzig.de

The manuscript was written through contributions of all authors. All authors have given approval to the final version of the manuscript. ‡These authors contributed equally.

KEYWORDS: Impedance Spectroscopy, ZnO Thin films, applied nanomaterials, glucose oxidase

ABSTRACT: Nanomaterials based on ZnO have been used to build glucose sensors due to its high isoelectric point, which is important when a protein like glucose oxidase (GOx) is attached to a surface. It also creates a biologically friendly environment to preserve the activity of the enzyme. In this work we study the electrical transport properties of ZnO thin films (TFs) and single crystals (SC) in contact with different solutions by using impedance spectroscopy. We have found that the composition of the liquid, by means of the charge of the ions, produces strong changes in the transport properties of the TF. The enzyme GOx and phosphate buffer solutions have the major effect in the conduction through the films, which can be explained by the entrapment of carriers at the grain boundaries of the TFs. These results can help to design a new concept in glucose biosensing.

# 1. Introduction

ZnO is a material that has important technological implications and has attracted special attention of researchers from different fields. It is a promising material for different applications, such as electronics, optoelectronics or piezoelectricity [1, 2]. But ZnO also attracted new attention for biosensor applications. A biosensor is an analytical device, which converts a signal from a molecular recognition element into an electrical signal, or other type of signal proportional to the concentration of a defined analyte. Nowadays, biosensors can be applied in a wide range of fields, such as medicine, pharmacology, food processing industries, environmental monitoring, defense and security, but most of this billionaire market is driven by medical diagnostics and particularly by sensors for glucose determination for people with diabetes [3]. With the advent of nanotechnology, biosensors have been improving in performance and versatility at a hectic pace. Advances in nanotechnology, MEMS, microfluidic chips and other very small-scale fabrication technologies will push progress in biosensors even further and faster [4].

The application of nanomaterials to the design of biosensors is desirable due to their high activity, good selectivity, and extraordinary specific surfaces. Nanostructured ZnO with its unique properties could provide a suitable substrate for the immobilization of enzymes, or biorecognition of other molecular elements, while retaining their biological activity. This leads to the construction of different biosensors with enhanced analytical performance [5]. Particularly, ZnO has been used to develop different types of sensors of, e.g., glucose, cholesterol, DNA, immunoglobulins, and $H_2O_2$, taking advantage of its biocompatibility, high stability, non-toxicity and the presence of native defects such as Zinc interstitials and oxygen vacancies [1]. It not only possesses a high surface area, but it also shows biomimetic and relatively high electron mobility features. It also presents a high isoelectric point (IEP) of about 9.5, which makes nanostructured ZnO materials suitable for adsorption of proteins with low IEPs, since protein immobilization is primarily driven by electrostatic interaction [5]. For example, ZnO nanowires were recently applied in the development of a piezoelectric nanoforce transducer for the detection of pathogens by binding antibodies to the biosensor surface [6].

Glucose biosensors have been intensively investigated due to their importance in the food industry and in the determination of glucose concentration in blood for when diagnosing and controlling people affected by diabetes. In the effort of developing more sensitive and reliable glucose biosensors, nanotechnology is being widely applied [5, 8]. Nevertheless ZnO nanostructures started to be used as glucose biosensors less than ten years ago [9], and since then an increasing number of research groups have reported the production of such biosensors by modifying electrodes with different ZnO nanostructures. Different analytes were used in the development of these biosensors, like the detection of peroxide, the most widely used analyte, the variation of pH [10], and the–non-enzymatic ZnO biosensors, just to name a few [11, 12]. In the case of pH detection, Lee and Chiu recently developed an AlGaN/GaN heterostructure and a ZnO-nanorod array to create a high-ion sensitive field-effect-transistor (ISFET) for the detection of glucose [10]. Nevertheless, the whole device is submerged in the sample solution. This method is not practical in the case of glucose determination in blood.

Another interesting application of ZnO nanostructures is in the non-enzymatic detection of glucose. In recent years there have been some interesting advances in this direction. The group of Tarlani et al. reported the shape controlled synthesis of ZnO nanostructures using a solvothermal process by adding different amino acids as bifunctional species with (or without) other additives, like urea or oxalic acid [11] to control the pH. These nanostructures have been supported on glassy carbon electrodes and tested concentrations of glucose between 1 to10 mM with an acceptable sensitivity response. Huh et al. reported hybrid 1D nanostructured Au/ZnO arrays for the development of an electrochemical and optical D(+)-glucose sensor [12]. This work takes into account the optical properties of the hybrid array, by using the concentration dependence of the plasmon absorbance to detect the D(+)-glucose, as a novel way for the quantitative analysis of the glucose in blood [12].

The use of ZnO thin films (TFs) is mainly widespread as gas sensors [1, 7, 13, 14]. On the other hand, the reported works on ZnO TFs as biosensors used this material submerging it in a liquid containing the desired analyte. Nevertheless, we believe that there are no published reports about the study of ZnO TFs in the way we propose here, where a drop of a solution is deposited on the TF. In this work we present a detailed study of the electrical response of ZnO thin films by impedance spectroscopy (IS) after dropping different liquid solutions, allowing us to understand how these aggregates affect the conduction paths in ZnO TFs. Changing the type and concentration of the solution, we investigate to what extent such a ZnO TF is applicable as a biosensor. This kind of measurements can be used, for example, when the liquid to be tested is blood, tear or saliva, thus reducing the sample volume and providing better applicableness. The results of our work indicate that the conduction through the grain boundaries is strongly affected by different aggregates and its contribution dominates the impedance value after the drying of the aggregate.

The paper is divided in two more sections. In the next section we describe all experimental details, from the preparation of the samples and the electrical contacts to the reagents used and the details of the impedance measurements. In Section 3 we present the results as follows. Firstly, in Sec. 3.1 we show the variability of the impedance for TF samples obtained from different regions of the same thin film. In Sec. 3.2 we present the impedance changes under visible light. In Sec. 3.3 the influence of a droplet of deionized water (DIW) on the impedance is discussed and we describe the general model we use to fit the results in terms of bulk and grain boundaries contributions. In Sec. 3.4 the impedance variation is discussed when a droplet of phosphate buffer solution (PBS) is placed on the ZnO TF. In Sections 3.5 and 3.6 a similar study is presented with a droplet of a high concentration of Glucose Oxidase (GOx) solution in PBS and a droplet of glucose 10 mM in PBS.

## 2. Experimental details

*Reagents*

Phosphate buffer (PBS) 0.07 M and pH 7.25 was prepared using $K_2HPO_4$ and $Na_2HPO_4 \cdot 7H_2O$. Glucose solution 1 M was prepared in the same buffer, and dilutions from the previous one at different concentrations (0.25, 0.75, 1.5, 3, 4, 5 and 10 mM). The GOx solution (Glucose Oxidase Type VII from Aspergilus Niger, Sigma-Aldrich) was prepared with concentrations of 48 mg/ml in PBS. All reagents were from Sigma-Aldrich, and DI water (18.2 MΩ) was used for all solutions.

*ZnO thin films*

ZnO TFs were grown using a PLD (pulsed laser deposition) system under a base vacuum pressure of $5 \times 10^{-6}$ mbar. c-Axis oriented ZnO films were grown on (100) sapphire (α-$Al_2O_3$) substrates using a Nd:YAG pulsed laser at 5 Hz, λ=248 nm, an energy of 650 mJ, a substrate temperature of 700°C and under an oxygen atmosphere of $125 \times 10^{-3}$ bar during the deposition. A ZnO target (99.999%) (Pi Kem LTD) was used. The substrate holder was placed 10 cm from the target and the film thickness was controlled by the number of pulses to obtain ~ 200 nm [15]. The deposition period was 15 minutes and after the temperature decreased to 25°C, each film was sonicated with a Bandelin Sonorex RK 100H sonicator with acetone, iso-propanol and DI water for 10 minutes. Finally, films were dried at 100°C during one hour. The single phase of the films was verified by using X-ray diffraction (Philips X'Pro wide angle X-Ray spectrometer), which spectrum is shown in Fig. 1. A Dual Beam Microscope - Nova NanoLab 200, FEI Company was used to characterize ZnO films with dried solutions on their surface.

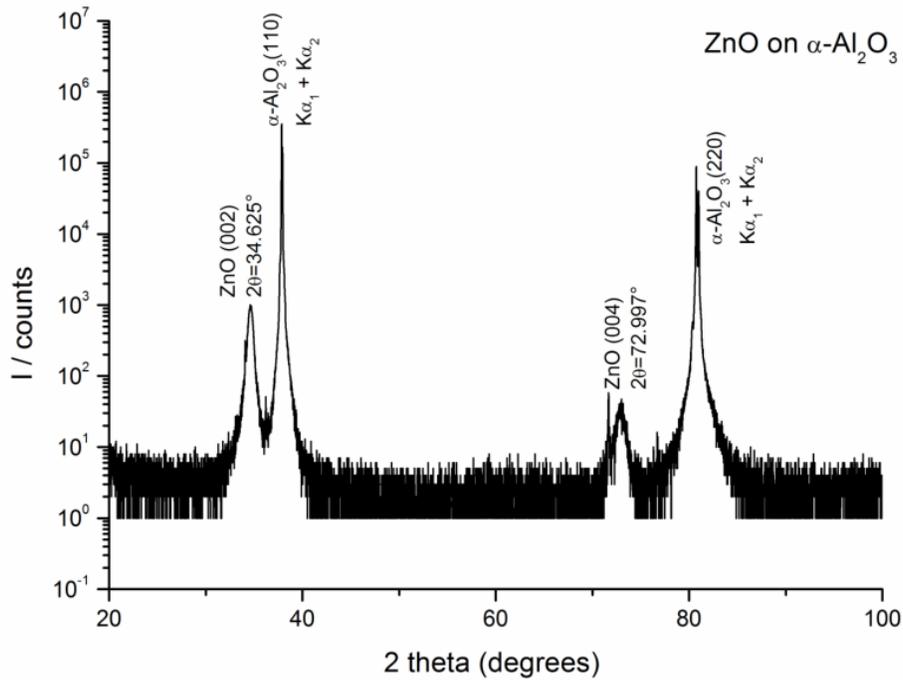

**Figure 1.** XRD spectrum of a ZnO thin film (sample TF01) with predominant c-axis orientation. The lattice parameter c for ZnO obtained from the spectrum is 5.18 Å.

The SEM image in Fig. 2 (plain view of the TF) shows the surface morphology of the used films. As seen, there is not open porosity to enable air inclusion or solution drain. The cross section SEM image (inset in Fig. 2) does not show porosity either (except for a large hole created during the film + substrate cutting procedure). Hence, the SEM images reinforce our proposed conduction model (grain + grain boundary).

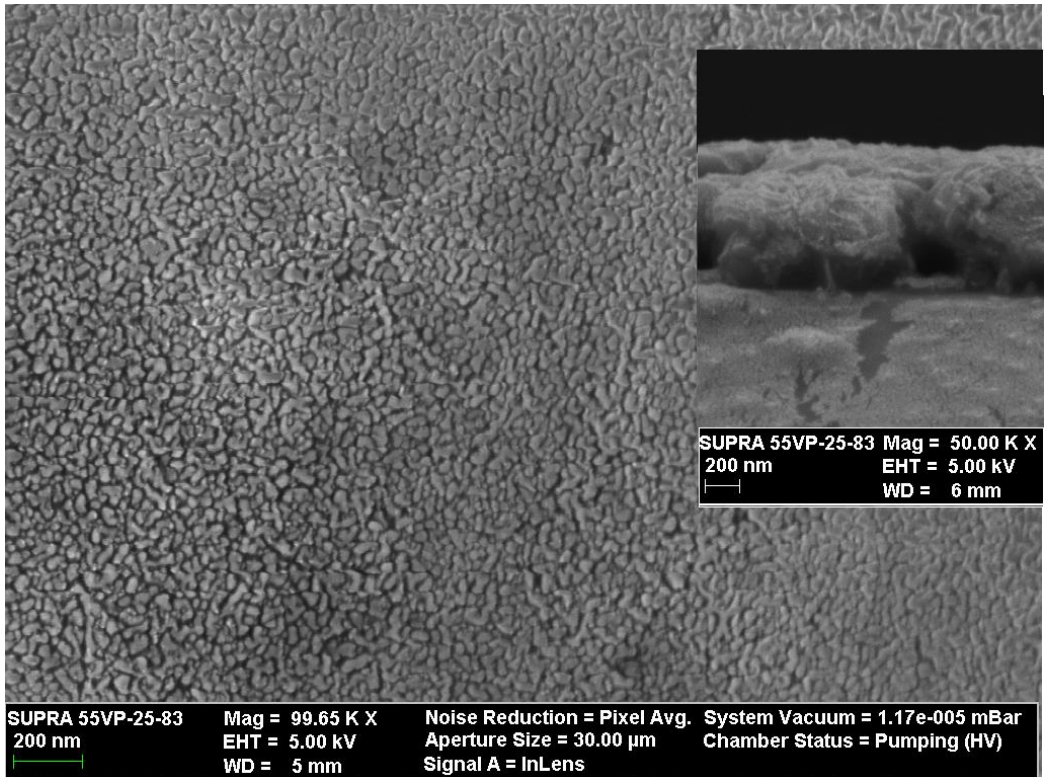

**Figure 2.** SEM images for plan-view of ZnO TF grown on C-plane sapphire substrates, and the cross-sectional view can be seen in the inset.

The observed structure also reflexes the grain and grain boundaries structure in agreement with the impedance spectroscopy measurements, which can change its electrical response as a function of the electrical charges contribution provided by the used solutions.

*Electrical Impedance measurements*

The electrical impedance was measured with frequency sweeps using an Impedance analyzer (Agilent 4294a). The signal was sinusoidal of 950 mV amplitude and a frequency range from 40 Hz to 3 MHz.

The thin film TF01 with dimensions of 5x5 mm$^2$ was contacted on their opposite extremes using a bipolar (two electrodes) configuration with the help of a Carl Zeiss Stemi 2000-C optical microscope. For comparison similar measurements were done with ZnO single crystals and are presented in Appendix B.

The electrical contacts on the ZnO samples were done using gold wires ($\Phi$ = 25 μm) attaching them with pure indium; the samples are mounted in a chip carrier Kyocera (Optocap, Scotland). A second batch of thin film samples, obtained from the sample TF2 grown in the same conditions as TF1, were cut with a Well Diamond wire saw in bars of 1 mm x 5 mm, as shown schematically in Fig. 3. In this way, we obtain 5 sub-samples grown under the same conditions, and were denoted as TF21 to TF25 to clarify that these bars were obtained from the same sample TF2.

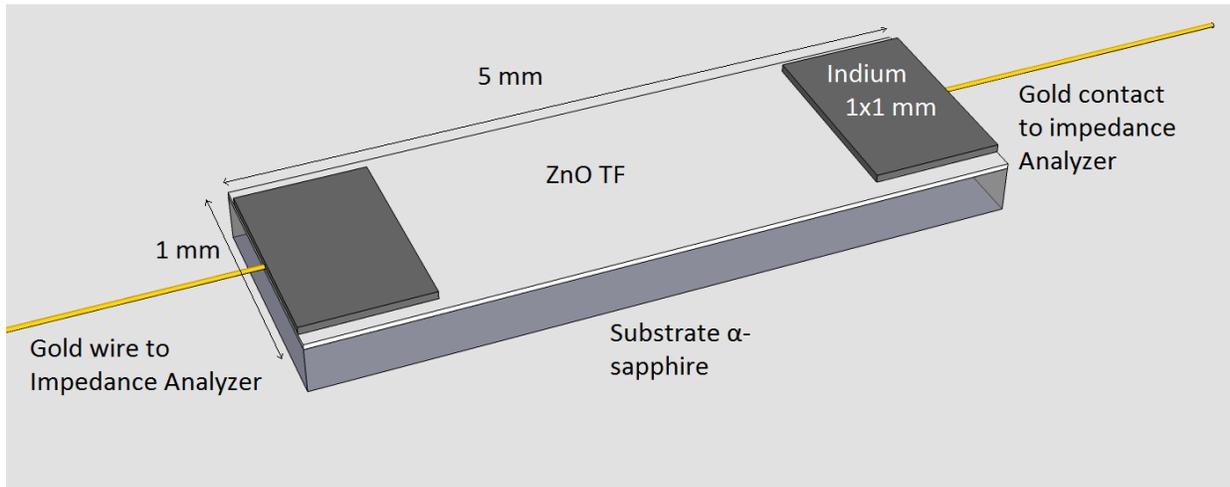

**Figure 3.** Diagram of the contacts on the surface of the film

Different solutions were added to the bars, as can be seen schematically in Fig. 4, and the electrical impedance was measured before the addition and every minute till the droplet of each solution was dried. To avoid residual contamination only one addition of each solution was made per sample.

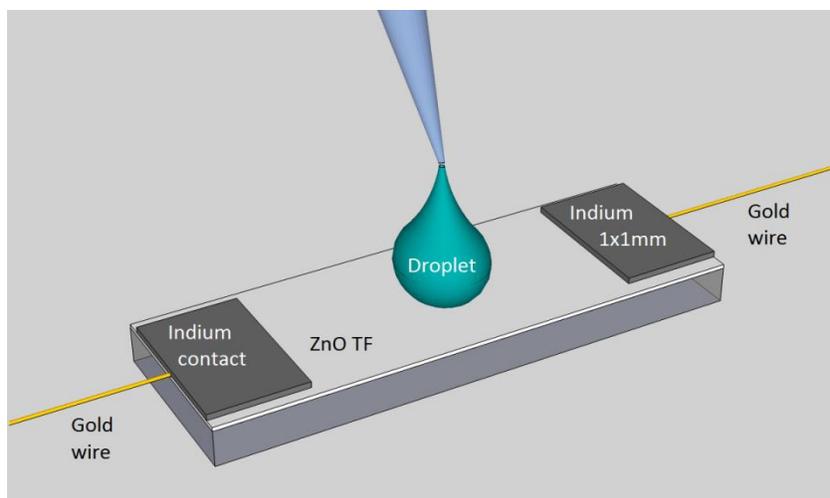

**Figure 4.** Sketch of the experimental procedure: The ZnO TFs were contacted with gold wires using indium. The electrical impedance was measured before and after dropping the droplet of different solutions. The droplet region was always of approximately 2 mm$^2$.

*Circuit models*

Using the software Zview 3.0 (© Scribner Associates, Inc.) circuit models were fitted to the impedance measurements. For native ZnO TF we can fit the system as a RC parallel circuit, as can be seen in Fig. 5. For the fits we used a constant phase element or CPE as a non-ideal capacitance instead of an ideal capacitor. It takes into account the frequency dispersion of the capacitance value. The CPE-P factor is equal to 1 when the element behaves as an ideal capacitor. Values higher than 0.8 indicate a closely capacitive behavior. For this sample the parameters that fit the data are shown in Table 1.

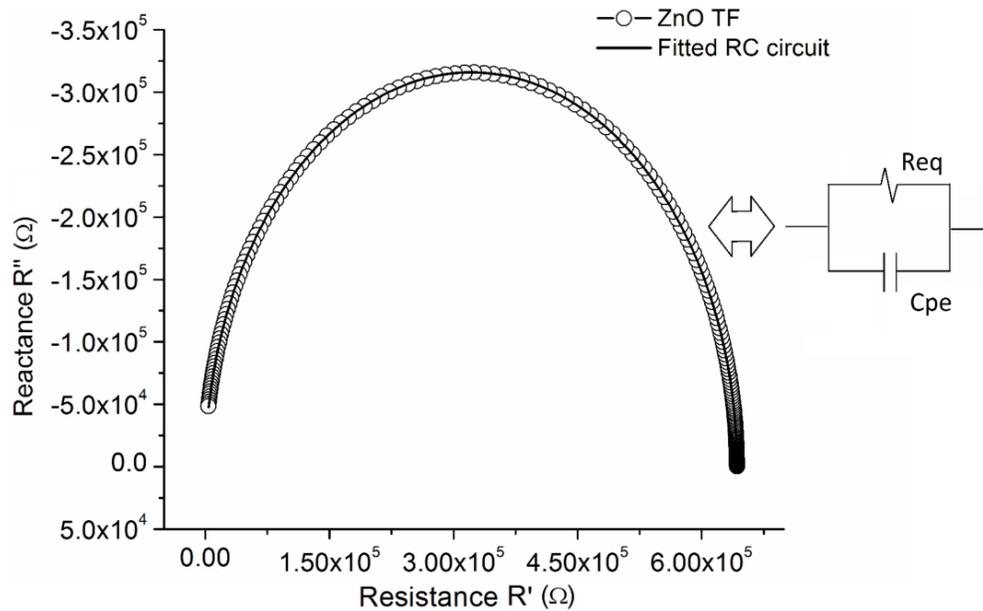

**Figure 5.** Typical Cole-Cole Plot for a ZnO TF measurement of the electrical impedance, and the corresponding electrical model of the conduction using the software Zview 3.0 ©. The experimental data the circles and the continuous lines the fittings. Typical values of the parameters R1 and CPE1 obtained from the fits are given in Table 1.

## 3. Results and Discussion

### 3.1 Film fabrication variability

Figure 6 shows the Cole-Cole plot of different ZnO TF-bars, denoted as TF21 to TF25, cut from the same TF. The electrical impedance was measured following the protocol described in section 2. Each measurement was performed in absence of light, and without the addition of any solution. The continuous lines inside experimental data are the fits obtained using the software Zview 3.0.

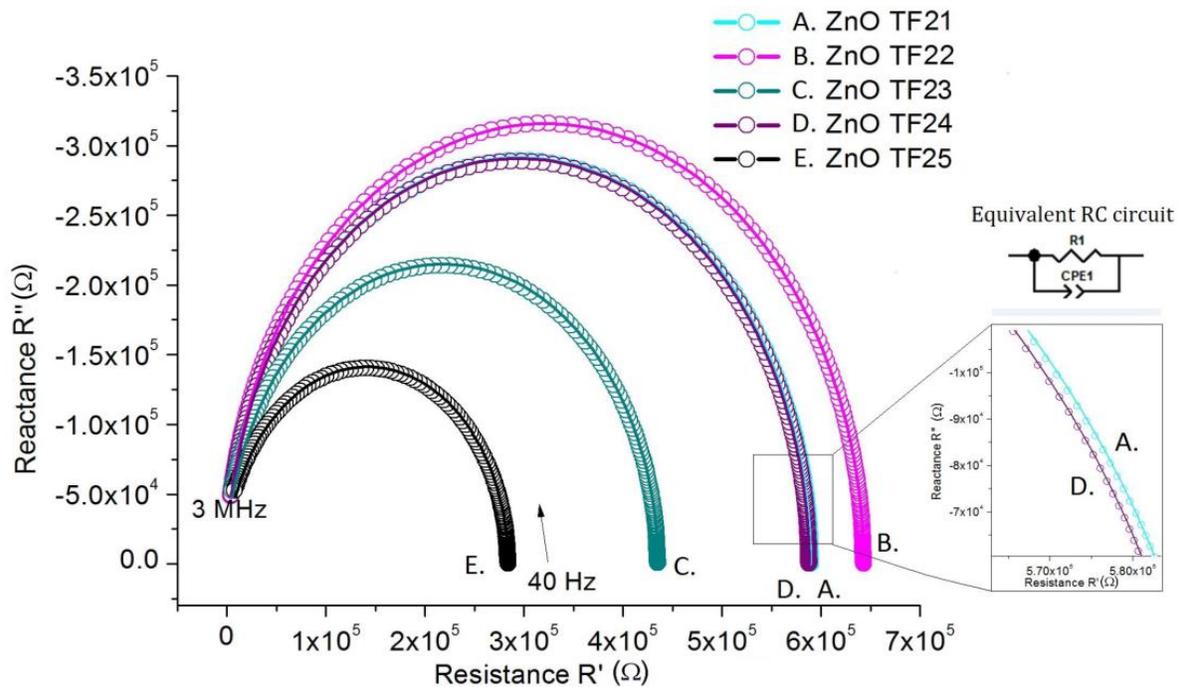

**Figure 6.** Impedance results (Cole-Cole representation) of five (TF21...TF25) ZnO TF-bars with the same geometry, obtained from the same TF.

We note that the impedance R' measured at the lowest frequency of 40 Hz shows an average variation of two, calculated for each TF (data not shown) obtained from the same substrate grown under the conditions described in the Experimental details Section. We shall see below that this variation of the impedance at low frequency is relatively small compared with the variation produced by certain liquids.

### 3.2 Changes in the electrical impedance due to ambient light

In order to explore changes due to the ambient light in the electrical impedance of ZnO TF, different experiments were carried out in the presence or absence of light. Firstly, the films were exposed to ambient light for one hour and then, the impedance was measured every minute in dark, see Fig. 7. We observe that when the sample was covered from light for 12.5 minutes the impedance increase was ~2.1%.

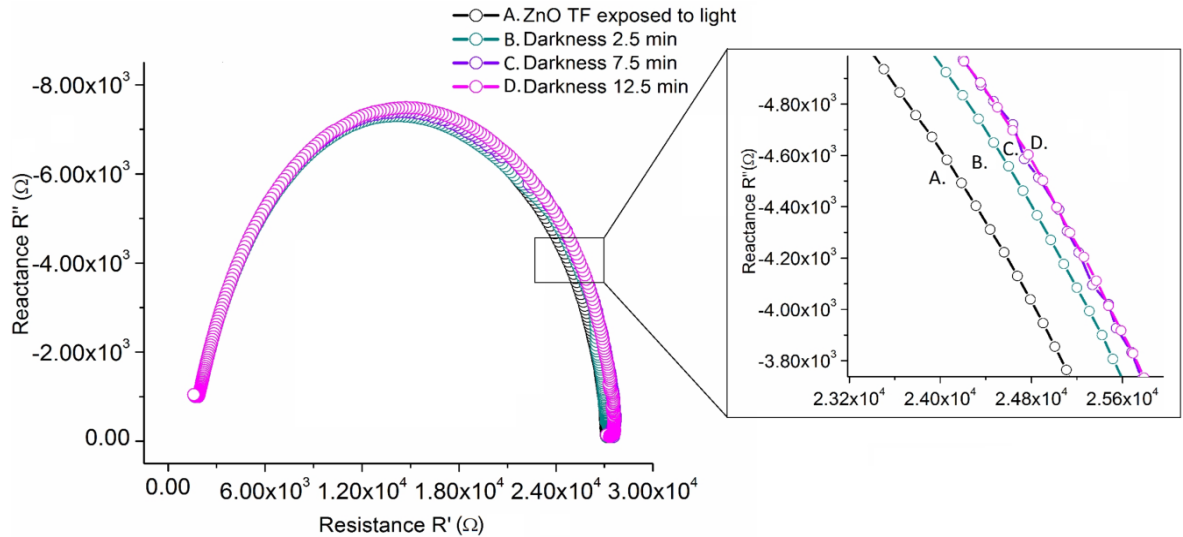

**Figure 7.** Cole-Cole plot of a ZnO TF at different times after turning off visible light. The TF was irradiated one hour with ambient light before starting the measurements. Impedance measurements were made every 30 seconds.

In the case of dark to light, the variation of the impedance in 12.5 minutes after exposing the sample to ambient light was 3.3%. In what follow, impedance changes of this order will be associated only to a light effect creating new charge carriers in the semiconducting film.

### 3.3 Deionized Water effect on ZnO thin films

In this section, we show the changes produced by a deionized water (DIW) droplet on the electrical impedance of the ZnO film. A droplet of 1 µL of DI water (18.2 MΩ, MilliQ Water © Millipore) was added at the center of the sample, see Fig. 8. The impedance was measured before and every 30 seconds after the addition of the droplet till it was dried.

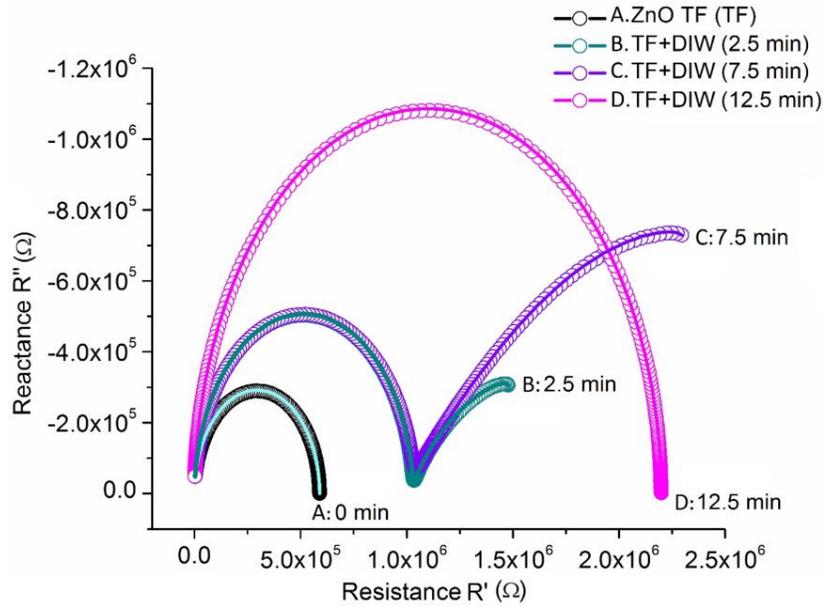

**Figure 8.** The Cole-Cole plot of the measurements are shown in the Figure, where only the measurements every 2.5 minutes are shown for clarity. We observe that the addition of a single water droplet produces large, time dependent changes in the impedance.

When DIW was added to the TF, the real part, as well as the module of the impedance |Z|, increases in the first minutes. After 12 hours the impedance values return partially to the initial state, see Figure A1 of Appendix A. To clarify this behavior and the reproducibility of the observed effect Figure A2 of Appendix A shows the module |Z| obtained at a constant frequency of 20 KHz, plotted as a function of time. The water droplet was added after two minutes. After 12 hours, when the droplet was completely dried, we have repeated the measurements. The results indicate a very good reproducibility of the observed behavior.

For a better interpretation of the impedance results shown in this and in the next sections, it is useful to model the samples as an array of grains with grain boundaries, as depicted schematically in Fig. 9, where a RC parallel circuit is used to fit the conduction of carriers through the grains and other circuit independently for each grain boundary [18]. We simplify the array of the all the contributions using the equivalent circuits shown in Table 1.

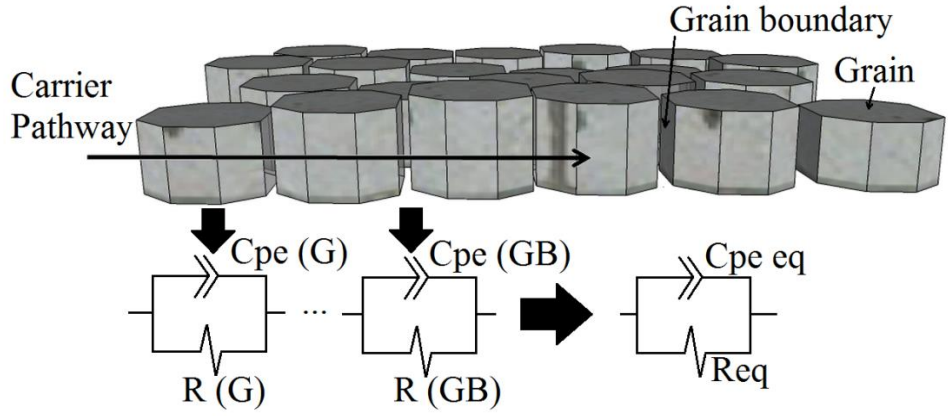

**Figure 9.** Assumed structure of the ZnO thin films Model of conduction with the different conduction pathways in the ZnO films. Each grain (G) is modeled with a parallel RC circuit, different from a similar circuit used for the grain boundaries (GB). The total conducting path is modeled as the equivalent of two circuits in series.

**Table 1.** Sample state, the equivalent circuit used to fit the impedance results, the values of the resistance $R_i$ (i = 1,2) and capacitances $Cpe_i$ (i = 1,2) obtained from the fits shown in Fig. 6.

| Sample state | Equivalent circuit | R1 | CPE1 | R2 | CPE2 |
|---|---|---|---|---|---|
| ZnO TF21 | R1–CPE1 | $5.89 \times 10^5 \, \Omega$ | $1.22 \times 10^{-12}$ F f=0.99 | - | - |
| 2.5 min | R1–CPE1, R2–CPE2 | $1.04 \times 10^6 \, \Omega$ | $1.42 \times 10^{-12}$ F f=0.99 | $6.43 \times 10^5 \, \Omega$ | $7.03 \times 10^{-9}$ F f=0.93 |
| 7.5 min |  | $1.04 \times 10^6 \, \Omega$ | $1.36 \times 10^{-12}$ F f=0.98 | $1.65 \times 10^6 \, \Omega$ | $4 \times 10^{-9}$ F f=0.85 |
| 12.5 min | R1–CPE1 | $2.2 \times 10^6 \, \Omega$ | $1.27 \times 10^{-12}$ F f=0.99 | - | - |

In general, values of capacitance in the range of $10^{-12}$ F are associated to grain conduction, whereas values between $10^{-9}$ to $10^{-10}$ F are associated to interface or grain boundaries conduction [16, 18]. From the fits we obtain that the droplet triggers an interface or grain boundaries contribution. The values obtained, see Table 1, for the capacitance are similar and or the order of $5\times10^{-9}$ F, whereas the resistance associated to this conduction path is around $10^6$ Ω. The corresponding values of resistance for the bulk conduction are similar. However, the capacitance values are much smaller suggesting that the remaining conduction path without droplet is mainly through the grains. Therefore, the fits indicate that the interfaces conduction path is mainly affected through the water molecules adding an extra resistance in the total conducting path of the carriers of the film. This increase in impedance would be associated to the capture of charge carriers by the water adsorbents.

For the ZnO TF with DIW we can plot (Fig. 10) the changes in resistance from Table 1 versus time. The equivalent resistance ($R_{EQ}$ DIW) is plotted when we have a single semi-circle in the impedance Cole-Cole plot, that is before the adding of DIW droplet and when the DIW droplet is dried.

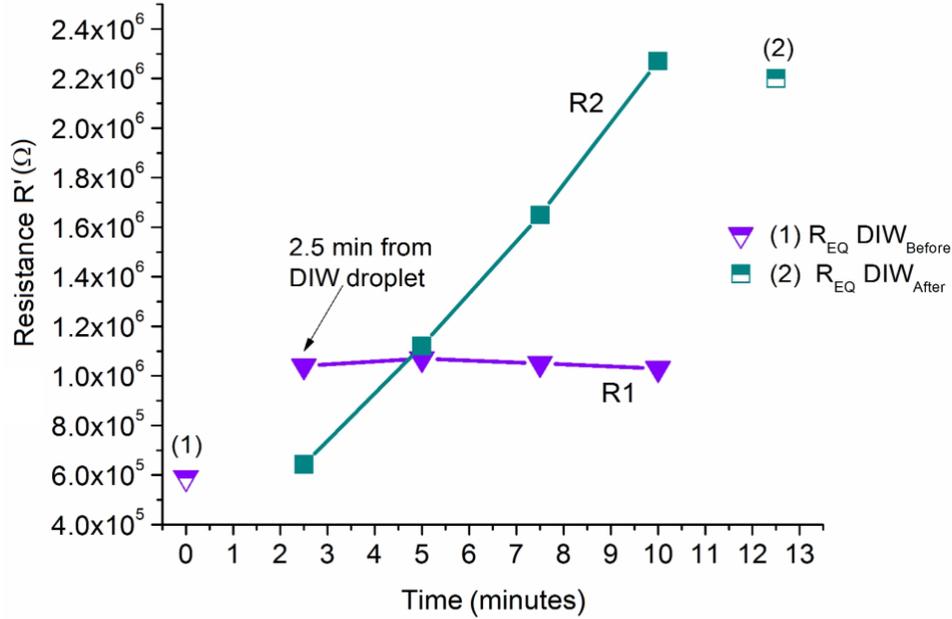

**Figure 10**. The parameters R1 and R2 obtained from the fits of the Cole-Cole plots of the sample TF021 vs time in minutes. The point Req DIW before at minute cero indicates the equivalent resistance before DIW and the point at 12.5 minutes indicates the equivalent resistance once the droplet of DIW was dried.

We observe that R1, associated to the conduction through the bulk, does not change with the addition of the DIW droplet, in contrast to R2. The observed behavior indicated that the transport properties are strongly influenced when DIW added due to the presence of grain boundaries, and in this case is a reversible process. DIW behaves as an impurity in the surface of the film [18].

**3.4 Phosphate buffer solution (PBS) on ZnO thin films**

The impedance measurements were first carried out in one ZnO TF bar, with a droplet of 1 μL of PBS solution on the surface of the film. The electrical impedance is shown in Fig. 11. The main difference with the results obtained after dropping a DIW droplet, is that in the present case, the impedance presents permanent changes due to solutes solved in water, see Fig. 8. Experimental data are shown in circles of different colors and the continuous lines through the data are calculated by fitting the circuits and parameters shown in Table 2.

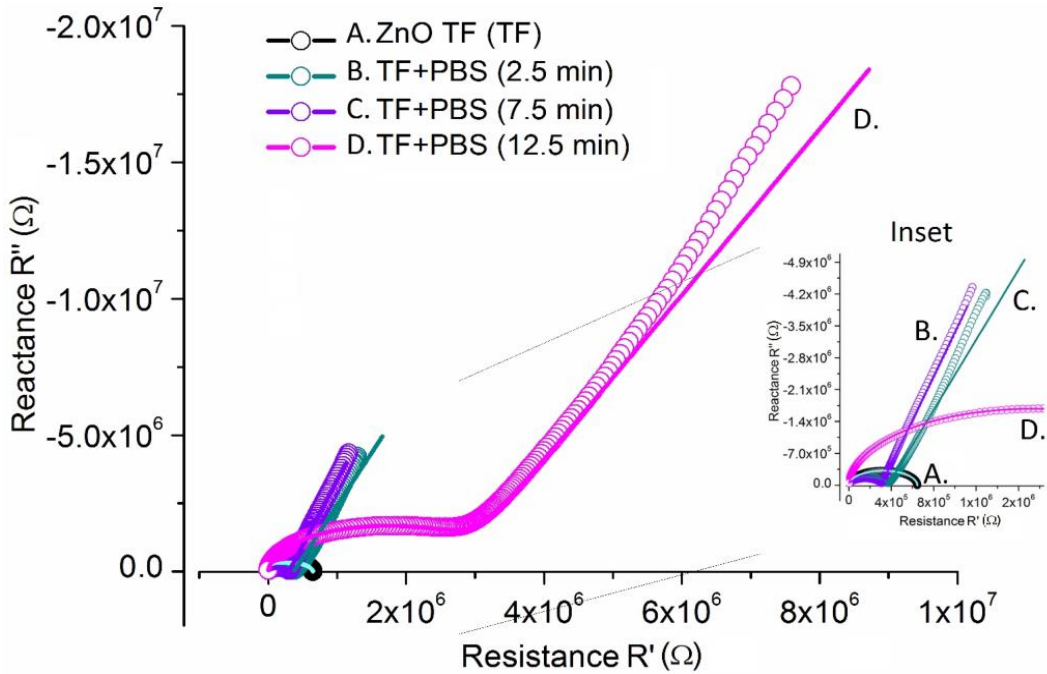

**Figure 11.** Cole-Cole Plot of the measured impedance for the sample TF22 with a droplet of buffer PBS. The data are shown every 2.5 minutes. The first plot (A. in black) showed the film without the solution. The inset at right shows the data at smaller values of the impedance in detail.

The system was fitted using different circuits based on the model described in the previous section. The equivalent electrical circuit for the ZnO TF before the PBS droplet is a parallel RC, see Table 1. While the PBS droplet is evaporating, the resistance associated to the conduction through the grain boundaries is very high. Therefore, we modeled this effect using only a CPE2 element, which can be interpreted as a capacitor with a small leaking current that represents an extremely high resistance in parallel with the capacitor. At low frequencies, the impedance of the TF with buffer solution raises two orders of magnitude. The fitting results are shown in Table 2. To clarify the effect of the buffer in ZnO TF, the modulus of the impedance is plotted at 20 KHz as a function of time, see Figure A3 in Appendix A.

**Table 2:** Equivalent circuits for PBS on ZnO TF22

| Sample state | R1 | CPE1 | CPE2 |
|---|---|---|---|
| ZnO TF22 | $6.45 \times 10^5 \, \Omega$ | $1.23 \times 10^{-12}$ F<br>f=0.99 | - |
| 2.5 min | $3.69 \times 10^5 \, \Omega$ | $1.50 \times 10^{-12}$ F<br>f=0.98 | $1.88 \times 10^{-9}$ F<br>f=0.84 |
| 7.5 min | $3.04 \times 10^5 \, \Omega$ | $1.35 \times 10^{-12}$ F<br>f=0.99 | $1.05 \times 10^{-9}$ F<br>f=0.80 |
| 12.5 min | $2.67 \times 10^6 \, \Omega$ | $1.26 \times 10^{-12}$ F<br>f=0.99 | $1.46 \times 10^{-10}$ F<br>f=0.80 |

We can see that the impedance grows to very high values after the droplet gets dry. Just after the addition of the droplet and in the first 12 minutes, the modulus of the impedance at 20 KHz decreases until the droplet begins to dry. This lower value of the impedance could be associated to the ionic content of the buffer solution. When the droplet gets dry, the salts of the buffer act as impurities in the grain boundaries, leading to a marked increase in the impedance value.

The work in Ref. [19] reported the use of different surfactants to assist the aqueous chemical growth of ZnO nanostructures. These surfactants, incorporated as impurities in the growth solution, determine the morphology and transport properties of the ZnO crystals. The authors stated that these impurities are localized on the surface and, in the case of the surfactant NaPTS, could behave as p-doping molecules, decreasing in this way the conductivity of the ZnO crystals [19]. These results support therefore the hypothesis that the salts of the buffer can be considered as impurities in the grain boundaries that increase the impedance value of the ZnO TF.

In Fig. 12 we show the obtained parameters R1, CPE1 and CPE2 vs. time, calculated in the equivalent circuits fitted from experimental data. The increase of R1 with the addition of the buffer is about one order of magnitude, while CPE1 remains constant and CPE2 decreases till the PBS droplet is dried. This last capacitance is related to the grain boundary conduction [17]. As can be seen in Fig. 12A, the value of resistance R1 at minute 12.5, is the equivalent resistance of the system when the droplet is dried.

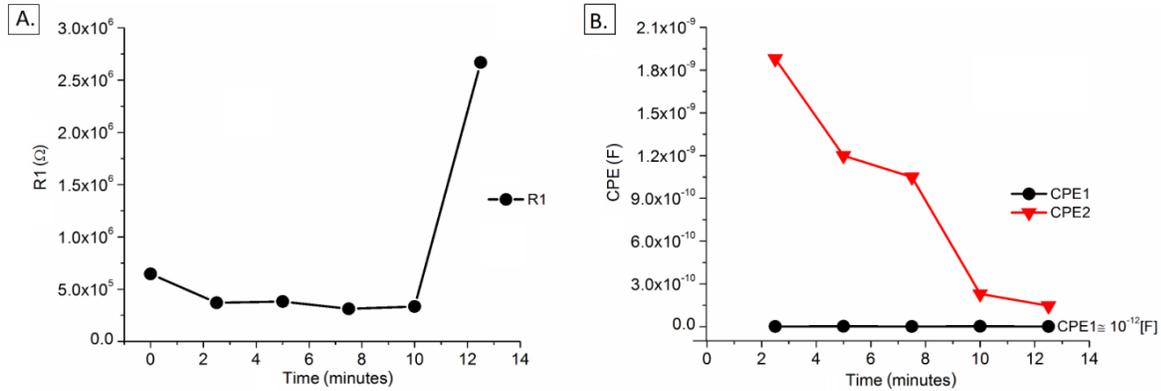

**Figure 12.** The parameters R1 (a), CPE1 and CPE2 (b) obtained from the fits to the Cole-Cole data of sample TF22 vs. time after leaving a droplet of the PBS solution.

The negative charge of the buffer ions, as well as the negative zones of the enzyme, can repel electrostatically the electrons flowing through the ZnO TF, generating a depletion zone between the grains. It is possible that the salts of the buffer act entrapping the electrons of the ZnO TF, which are the major carries in the semiconducting film.

As it is known, in ZnO films, two conduction paths exist; one path is through the bulk of each monocrystal and the other through the grain boundaries [20]. Therefore, the transport properties of the TF are clearly affected by the presence of impurities at the grain boundaries.

In order to corroborate the influence of the grains and their grain boundaries on the frequency dependent conductivity of the ZnO TF, we have measured the impedance of commercial ZnO single crystalline substrates (5 x 5 x 1 mm$^3$, Crystec GmbH) grown by hydrothermal method and using both faces, the oxygen and Zn terminated ones. These results and the comparison are given in an Appendix B at the end of the paper.

## 3.5 Glucose Oxidase (GOx) in PBS on ZnO TF

The concentration of Glucose oxidase used for these experiments is 48 mg of enzyme per ml of buffer PBS.

A droplet of 1 µL of GOx enzyme in high concentration was added to a ZnO TF, and the impedance measurements were made every 2.5 minutes, as in the previous cases. The solution behaves similar to the buffer, but the droplet dries faster than the buffer alone because of the high concentration of enzyme. The impedance results are shown in Fig. 13.

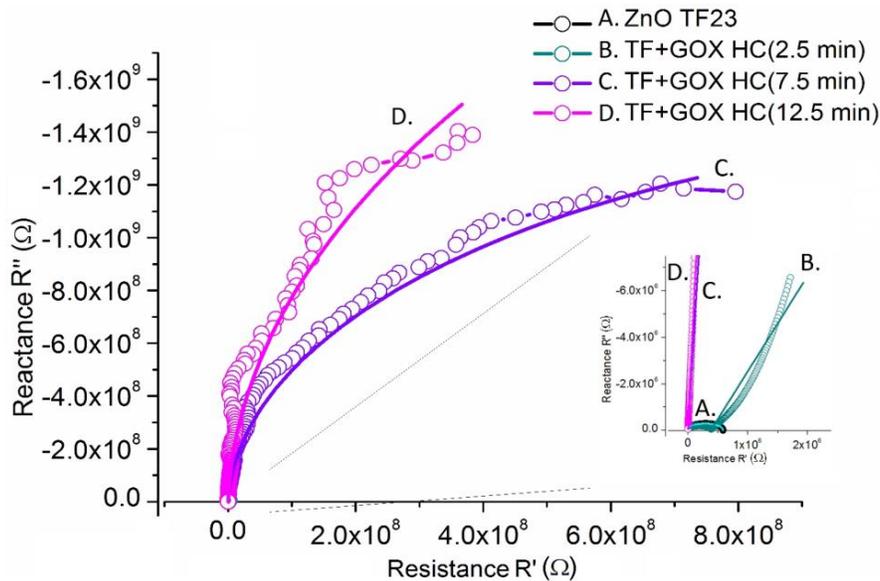

**Figure 13.** Cole-Cole plots for a ZnO TF with a droplet of enzyme GOx in high concentration in PBS. The solution was dried before 12.5 minutes. The inset shows the results for the film without the droplet of GOx and 2.5 minutes after dropping the droplet. The continuous lines are the fits to the data.

The parameters of the equivalent circuits that fit the experimental data of Fig. 13 are given in Table 3. We observe changes in the resistance of four orders of magnitude, see Fig. 14.

**Table 3.** Equivalent circuits used to fit the impedance data for the ZnO TF with a droplet of GOx

| Sample state | R1 | CPE1 | CPE2 |
|---|---|---|---|
| ZnO TF23 | 5.88x10⁵ Ω | 1.21x10⁻¹² F<br>f=0.99 | - |
| 2.5 min | 3.83x10⁵ Ω | 1.23x10⁻¹² F<br>f=0.99 | 7.8x10⁻¹⁰ F<br>f=0.85 |
| 7.5 min | 3.17x10⁹ Ω | 1.25x10⁻¹² F<br>f=0.99 | - |
| 12.5 min | 7.54x10⁹ Ω | 1. 2x10⁻¹² F<br>f=0.99 | - |

In Figure A4 in the Appendix A we show the effect of the enzyme solution after the addition of the GOx solution for two different samples (ZnO TF23 and TF24), showing the same behavior of the modulus of impedance at 20 KHz vs time. It can be seen that the impedance increases up to 2600% referred to the initial modulus of the film without any solution. After 12 hours the impedance of the ZnO TF remains as high as the day before when the enzyme solution was dried. This can be attributed to the high affinity between the enzyme GOx and ZnO at neutral pH. The high isoelectric point (IEP) of ZnO facilitates the strong electrostatic interaction with the GOx enzyme, which has a low IEP of 5.4.

Figure 14 resumes the changes of the parameter R1 after the addition of the different solutions (DIW, PBS Buffer and GOx of high concentration). We attribute changes in resistance, in the first minutes, to small leaking of ionic current due to salts solved in PBS and in the enzyme solution. Only in the case of DIW (De-ionized water) we attribute the impedance increase to oxygen molecules solved in water, which acts as impurities on ZnO TF's surface, increasing the impedance value for several hours. This is the reason why the value of the impedance, only in this case, is almost restored after 12 hours when the experiment was repeated with DIW in the same sample TF21. This data is shown in Figure A1 of Appendix A.

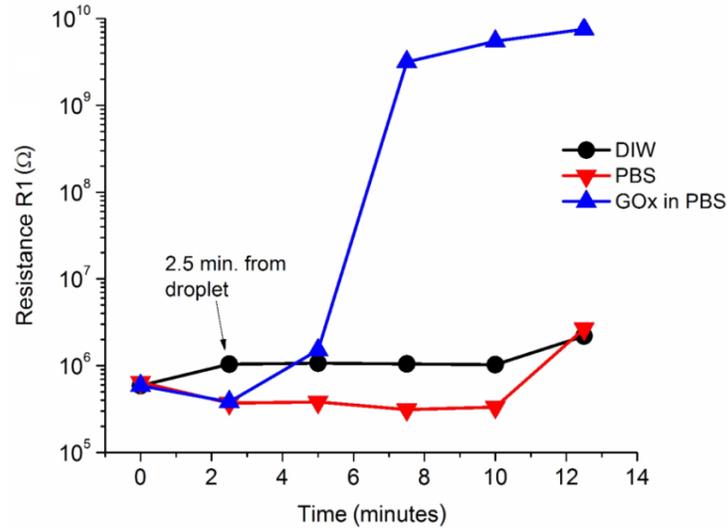

**Figure 14.** The parameter R1 vs. time obtained from the fits to the IS results for the TF after dropping droplets of DIW, buffer (PBS) and GOx with a concentration of 48 mg/ml.

The parameter R1 shows the same behavior as the one obtained for the impedance modulus, see Fig. 14. We have observed that after the drying of the droplet there is some kind of cracks of the enzyme layer over the TF. These cracks may lead the salts crystals to be in contact with the surface of the TF producing the increment of impurities at the grain boundaries, increasing in this way the measured impedance. Figure 15 shows SEM images of the TF with the dried enzyme solution. Crystalline structures on the TF can be attributed to the buffer salts once the enzyme solution is dried on the surface.

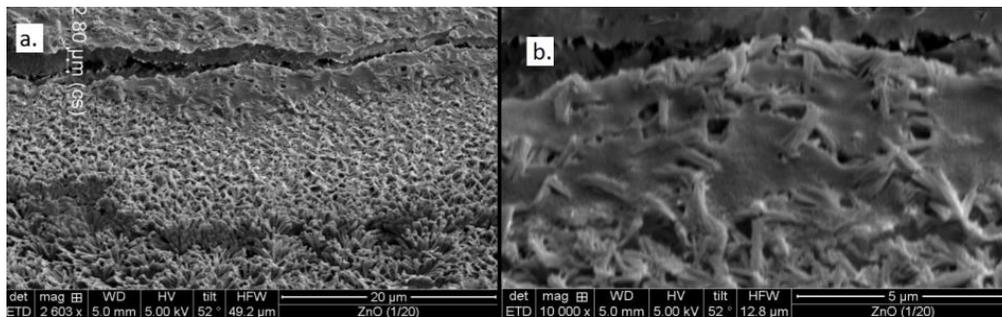

**Figure 15.** SEM images of a ZnO TF with the enzyme GOx at 48 mg/ml in PBS with magnification of 2600x (a.) and 10.000x (b.)

To understand the large change in R1 we take into account the following points. The GOx enzyme has a high content of acidic amino acids, which contribute to the net negative charge at neutral pH [21]. Then, we can provide two possible explanations for the behavior of the TF with the addition of the GOx solution. The first possibility is that, due to the high negative charged enzyme, a depletion zone is generated between the TF grain boundaries, leading to a huge increase in impedance. The other possibility is that the ions of the solution as well as the enzyme trap the electrons from the surface of the TF.

**3.6 Glucose in PBS on ZnO TF**

Different glucose solutions in PBS buffer were dropped on the ZnO TF bars. The effect of glucose was measured and modeled using an RC parallel circuit and a double RC as in the case of DIW.

Glucose at high concentration of 10 mM in PBS has a different effect on the conduction on the ZnO TF25 than the others solutions.

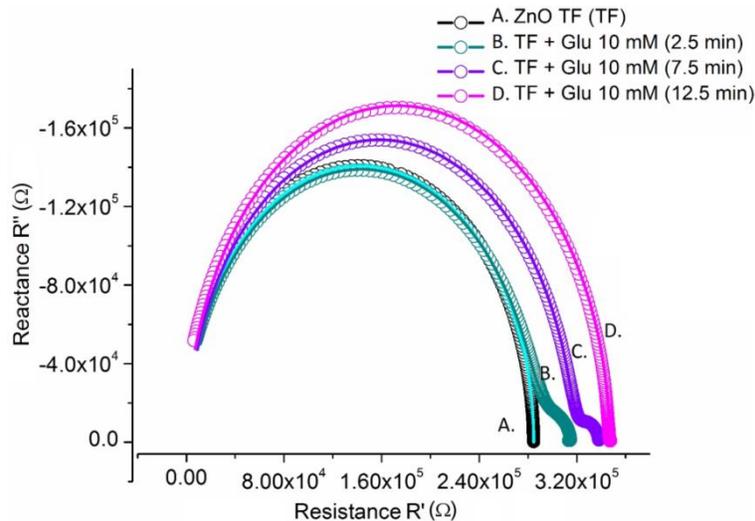

**Figure 16.** Glucose 10 mM in PBS on ZnO film, data is showed every 2.5 minutes after adding the droplet of glucose of 10mM in PBS. Continuous lines are the fitted data. The inset plot shows the fittings at lower frequencies.

The parameters of the fitting circuits to the data shown in Fig. 16 are resumed in the following Table 4. We observe that the changes in the impedance are relatively small after dropping the glucose droplet. A possible reason for this result could be that the glucose molecules do not have charges and shield in part the effects of water and of the buffer salts.

**Table 4.** Equivalent circuits used to fit the impedance data for the ZnO TF with a droplet of glucose 10 mM

| Sample state | R1 | CPE1 | R2 | CPE2 |
|---|---|---|---|---|
| ZnO TF25 | $2.85 \times 10^5 \, \Omega$ | $1.2 \times 10^{-12}$ F $f=0.993$ | - | - |
| 2.5 min | $2.89 \times 10^5 \, \Omega$ | $2.3 \times 10^{-12}$ F $f=0.95$ | $25.9 \times 10^3 \, \Omega$ | $4.2 \times 10^{-9}$ F $f=0.85$ |
| 7.5 min | $3.18 \times 10^5 \, \Omega$ | $2.2 \times 10^{-12}$ F $f=0.95$ | $20.6 \times 10^3 \, \Omega$ | $1.6 \times 10^{-8}$ F $f=0.83$ |
| 12.5 min | $3.47 \times 10^5 \, \Omega$ | $1.24 \times 10^{-12}$ F $f=0.99$ | - | - |

*%errors in the elements are lower than 5% for this table

With glucose solutions at lower concentrations we obtain similar results as with the buffer solution, as expected actually. The modulus of the impedance grows up several orders of magnitude ($10^4$ to $10^6$) after the droplet dries.

**Conclusions:**

This paper presented a detailed study of the electrical response of ZnO thin films by impedance spectroscopy after dropping different liquid solutions, and how these aggregates affect the conduction paths in ZnO thin films. Changing the type and concentration of the solution, we investigate to what extent such a ZnO thin film is of interest for biosensing applications. The results of our work show that the grain boundaries conduction is strongly affected by the different aggregates and its contribution dominates the impedance value after the drying of the aggregate. Equivalent RC circuits were used to fit the data. The largest increase in impedance is obtained after the addition of the enzyme Glucose Oxidase at high concentration while for 10 mM glucose addition impedance varies in small amount.

For the development of non-invasive glucose sensors, the analyte should be measured in tears, saliva or sweat instead of blood. Since the concentration of glucose is much lower in these fluids, the challenge is to design sensors with very high sensitivity and also with the capacity to detect them in small volumes. For the application in glucose sensing the concentration of GOx should be adjusted in order to have a monolayer on the surface of the film. Since those changes in the conductivity of the film are permanent, and the enzyme is covalently attached to the ZnO surface, the new information achieved in our study helps to understand the interaction of molecules with a nanostructured oxide film that can be used in the design of glucose biosensors.

**Acknowledgment**

The support of the National Scientific and Technical Research Council – Argentina (CONICET), Ministry of Science, Technology and Productive Innovation (MinCyT) and the *Bundesministerium für Bildung und Forschung* (BMBF) is gratefully acknowledged.

**Appendix A**

Additional Figures that complement the measurements are shown in this supporting file.

**Appendix B**

Comparison of the electrochemical impedance spectroscopy between a commercial ZnO single crystal and ZnO thin film with PBS droplets.

# Appendix A

# Nanostructured ZnO films: a study of molecular influence on transport properties by impedance spectroscopy

Luciano D. Sappia [a,b], Matias R. Trujillo [a,b], Israel Lorite [d], *Rossana E. Madrid [a,b], Monica Tirado [c], David Comedi [c], **Pablo Esquinazi [d]

[a] Instituto Superior de Investigaciones Biológicas (INSIBIO), CONICET, Chacabuco 461, T4000ILI – San Miguel de Tucumán, Argentina.

[b] Laboratorio de Medios e Interfases (LAMEIN), Departamento de Bioingeniería, Fac. de Cs. Exactas y Tecnología, Universidad Nacional de Tucumán, Av. Independencia 1800, 4000 – San Miguel de Tucumán, Argentina.

[c] Laboratorio de Nanomateriales y Propiedades dieléctricas de la Materia, Departamento de Física, Universidad Nacional de Tucumán, Avenida Independencia 1800, Tucumán, Argentina.

[d] Superconductivity and Magnetism Division – University of Leipzig - Leipzig, Germany.

---

*Corresponding author: Tel. +54 381 4364120. E-mail: rmadrid@herrera.unt.edu.ar

** Corresponding author: Tel. +341 97-32750. E-mail: esquin@physik.uni-leipzig.de

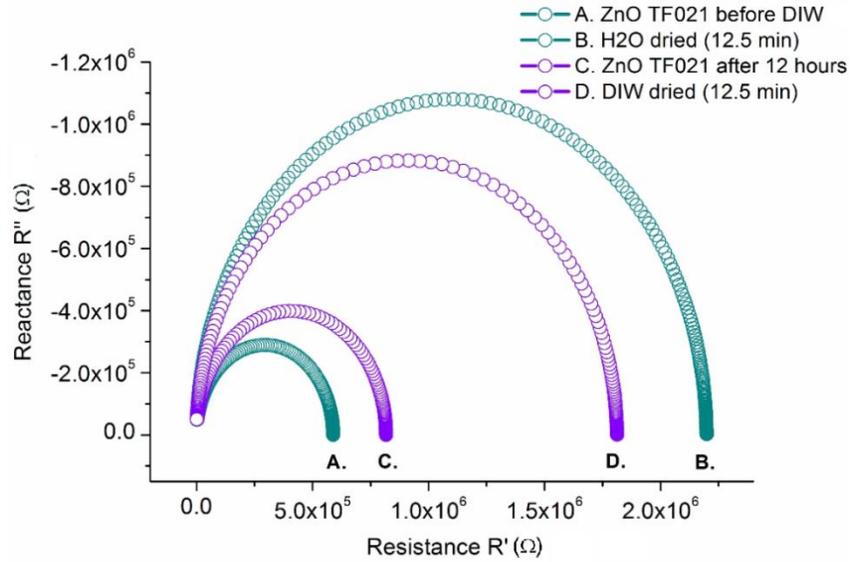

**Figure A1.** Cole-Cole plots of the TF S021 sample before and 12.5 minutes after dropping the droplet of water, curves (1) and (2), respectively. Curve (3) represents the results obtained after 12 hours and curve (4) was measured 12.5 minutes after dropping a new water droplet on the same place.

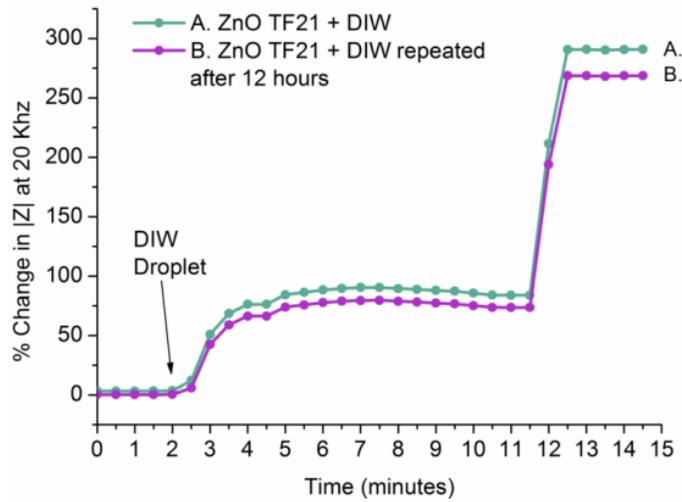

**Figure A2.** Relative change of the impedance module |Z| vs. time at a fixed frequency of 20 kHz, before and after adding 1 µL of a DIW droplet at the middle of the sample S021 at the minute 2. The film showed the same behavior when the experiment was repeated after 12 hours.

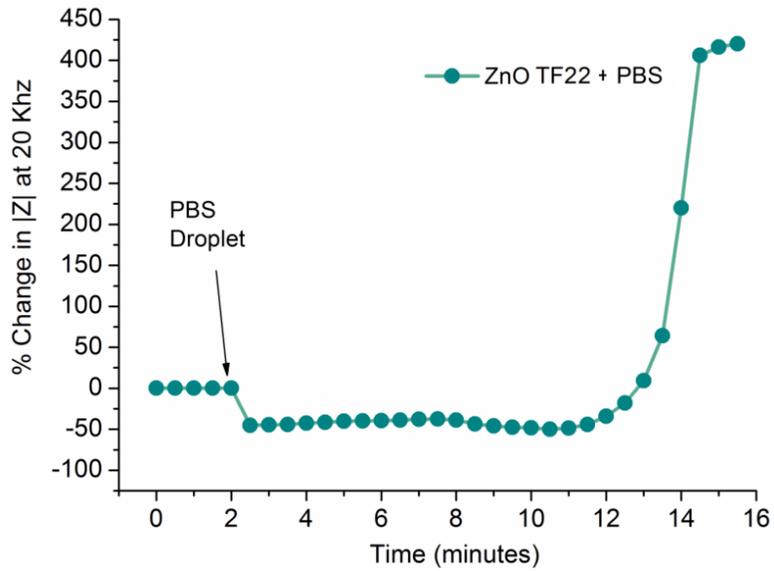

**Figure A3.** Relative change (in %) of the modulus of the impedance for ZnO TF at 20 KHz. The PBS droplet was added at minute 2.

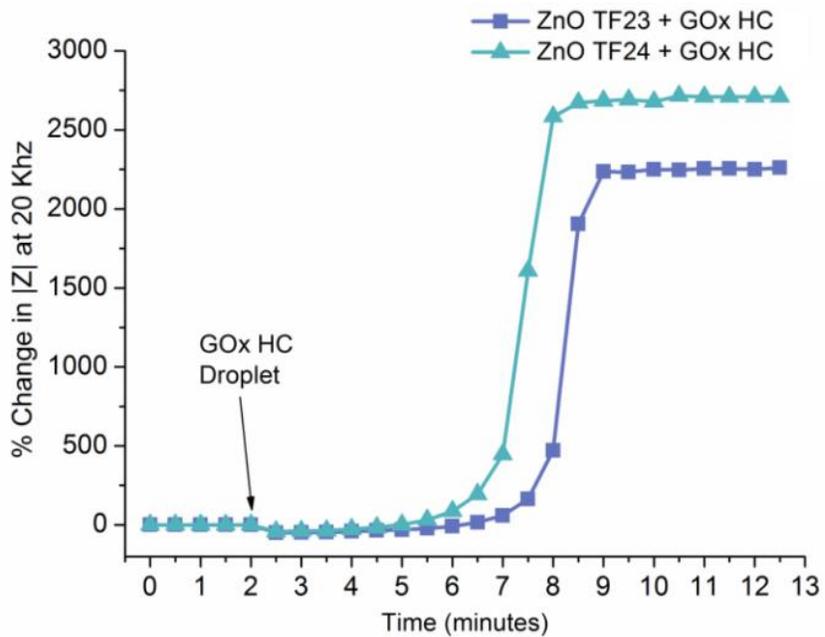

**Figure A4.** Cole-Cole Plot for a ZnO TF with a droplet of enzyme GOx in high concentration in PBS, repeated in two different samples TF023 and TF024

# Appendix B

# Nanostructured ZnO films: a study of molecular influence on transport properties by impedance spectroscopy

Luciano D. Sappia [a,b], Matias R. Trujillo [a,b], Israel Lorite [d], *Rossana E. Madrid [a,b], Monica Tirado [c], David Comedi [c], **Pablo Esquinazi [d]

[a] Instituto Superior de Investigaciones Biológicas (INSIBIO), CONICET, Chacabuco 461, T4000ILI – San Miguel de Tucumán, Argentina.

[b] Laboratorio de Medios e Interfases (LAMEIN), Departamento de Bioingeniería, Fac. de Cs. Exactas y Tecnología, Universidad Nacional de Tucumán, Av. Independencia 1800, 4000 – San Miguel de Tucumán, Argentina.

[c] Laboratorio de Nanomateriales y Propiedades dieléctricas de la Materia, Departamento de Física, Universidad Nacional de Tucumán, Avenida Independencia 1800, Tucumán, Argentina.

[d] Superconductivity and Magnetism Division – University of Leipzig - Leipzig, Germany.

---

*Corresponding author: Tel. +54 381 4364120. E-mail: rmadrid@herrera.unt.edu.ar

** Corresponding author: Tel. +341 97-32750. E-mail: esquin@physik.uni-leipzig.de

**Comparison between single crystal and thin film with PBS:**

In order to corroborate the influence of the grains and their grain boundaries on the frequency dependent conductivity of the ZnO TF, we have measured the impedance of commercial ZnO single crystalline substrates (5 x 5 mm$^2$) using both faces, the oxygen and Zn terminated ones. For comparison, measurements on ZnO TF were done using the whole substrate. The contacts were made in the opposite vertex of the film or single crystal and the droplet was added at the center, see Fig. B1. Because the large changes in the impedance were measured when the droplet was buffer PBS, we compare the impedance results for this case. The volume of the droplet was 2 µL.

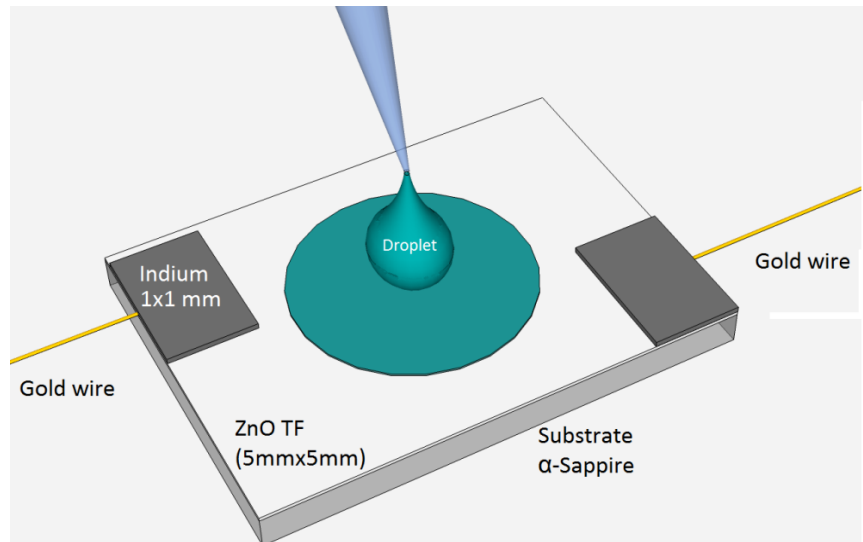

**Figure B1.** Scheme of the experiment used to compare the effect of the buffer between the thin film and the single crystals. The same geometry and electrode arrangement were used for both, the ZnO TF and single crystalline substrate.

Figure B2 shows the results for the TF (a) and for both faces of the single crystalline ZnO substrate (b, c). One clearly recognizes the large changes in the impedance of the TF after dropping the PBS droplet, while, in comparison, the measured changes in the single crystal are negligible.

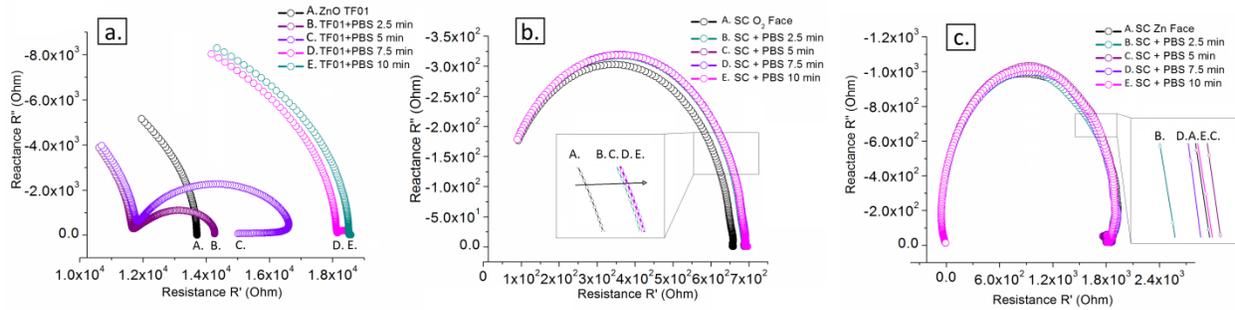

**Figure B2.** Cole-Cole Plots for the ZnO TF (a), ZnO single crystal O-face (b) and a ZnO single crystal Zn-face (c) after adding of 2 µL of buffer PBS at the center of the samples, at different times.